# Broadband Supercontinuum Generation in PCF, HNLF and ZBLAN Fiber with a Carbon-Nanotube-based Passively Mode-locked Erbium-doped Fiber Laser

*Yemineni Sivasankara Rao,[1] W. J. Lai,[1,2] A. Alphones[1] and P. P. Shum[1],

[1]*School of Electrical and Electronic Engineering, Nanyang Technological University, Singapore-639798.*
[2]*Temasek Laboratories @ NTU, Nanyang Technological University, Singapore- 637553.*
*sivasank001@e.ntu.edu.sg

**Abstract:** We demonstrate the broadband supercontinuum (SC) generation in photonic crystal fiber (PCF), highly nonlinear fiber (HNLF) and ZBLAN (ZrF4-BaF2-LaF3-AlF3-NaF) fiber with a passively mode-locked erbium-doped fiber laser (EDFL). The passively mode-locked EDFL incorporates a CNT-based saturable absorber and has achieved a pulse width of 620 fs with a pulse repetition rate of 18 MHz. The spectral broadening phenomena inside each fiber has been observed with respect to the variation in seed pulse power. The SC spectrum bandwidth of 1050 nm, 1400 nm, and 2000 nm has been achieved using PCF, HNLF, and ZBLAN fiber respectively.

## 1. Introduction

Supercontinuum (SC), is a process of generating broadband optical spectrum by launching optical pulses into a nonlinear medium, where the continuous interaction of laser pulses with nonlinear optical medium leads to the emission of a wider optical spectrum, which has a bandwidth many times greater than that of the launched pulses. The key nonlinear physical phenomena involved in this processes are stimulated Raman scattering (SRS), modulation instability (MI), cross-phase modulation (XPM), self-phase modulation (SPM), four-wave mixing (FWM) and soliton related dynamics. The first demonstration of SC generation in bulk media was reported by Alfano and Shapiro in 1970 [1-2]. Since then, this research field remains interesting for many researchers and it has enormous attractive application potential in the field of optical coherence tomography [3], micromachining [4], nonlinear frequency conversion [5], and light detection and ranging (LIDAR) [6].

The SC generation strength depends on the nonlinearity length of the medium and the interaction length of the seed laser with the medium. Therefore, optical fiber media is preferred for flat and broadband SC generation, since it provides efficient seed laser beam confinement in a small fiber core over a long distance. Moreover, fiber-based SC sources are attractive, for their great reliability, stability, and compact design. Realization of SC sources from all-fiber-based design consists of two essential parts: a nonlinear medium and a seed laser. Furthermore, to achieve an SC spectrum with a high output power, a cascaded amplifying system will be added after the seed laser. There are a large variety of fibers that can be used as nonlinear media such as single mode fiber (SMF), silica-based highly nonlinear fiber (HNLF), photonic crystal fiber (PCF), and ZBLAN (ZrF4-BaF2-LaF3-AlF3-NaF) fiber, etc. The seed laser can generate femtosecond (short) pulses, picosecond or nanosecond (long) pulses or even continuous-wave (CW) light, which are then amplified to reach the required power level. It is worth noting that the SC generation using CW laser as the seed has the advantages of lower cost, higher stability, and simpler configuration when compared with that generated using pulsed seed lasers. However, the spectral broadening efficiency of a CW laser is on the much lower side than in the case of pulsed lasers, which makes use of long length nonlinear fibers a necessity. Moreover, achieving the high output power with the CW laser is quite challenging. Therefore, pulsed lasers are preferred as pump sources to achieve high SC output power and with a short fiber length [7].

The flatness and bandwidth of the generated SC spectrum are related to the duration of pump pulse and the relationship between the pump wavelength and the zero-dispersion wavelength (ZDW) of the nonlinear medium [8]. Depending on the pump duration and in which dispersion region (normal or anomalous) the fiber is pumped, the dominant nonlinear phenomena responsible for the spectral broadening change accordingly and so does the SC spectrum dynamics. Suppose the fiber is pumped with long pulses in normal dispersion region, the spectral broadening and SC dynamics is dominantly decided by SRS and FWM, while when it comes to short pulses, SPM is the dominant process [9]. On other hand, in anomalous dispersion region, MI is the main nonlinear phenomena in case of long pulses [9], and soliton fission [10] and soliton self-frequency shift [11] is responsible for spectrum broadening in the case of short pulses.

To offer the possibility of broad and flat SC spectrum generation from a nonlinear fiber, efficient activation of nonlinear properties is much needed. Such nonlinear properties activation is possible with the high input power. However, achieving a high instantaneous power using a CW laser is much more challenging as compared to using a pulsed laser. Therefore, pulsed lasers obtained by applying the mode-locking technique to CW lasers. Mode locking is a technique in which the phase between neighbouring longitudinal modes is matched by imposing periodic loss over the CW laser signal in the cavity in each roundtrip. This kind of phase locking can be obtained by active or passive methods. Each of these mode-locking techniques has its unique way of producing optical pulses with different attractive properties. Through active mode-locking techniques, optical pulses with high pulse quality and high repetition rates can be generated. However, pulse duration and pulse peak power are limited. On other hand, higher peak power and shorter pulse width can often be achieved through passive mode-locking techniques, and the peak power can be boosted to over 1 MW after some amplification [12-13]. Such high peak power and short pulse width systems are attractive in nonlinear optics to generate SC [14]. There are various passive mode-locking techniques available to date, such as nonlinear polarization rotation (NPR) [16], nonlinear amplifying loop mirrors (NALM) [15], and saturable absorbers (SA) [17-18]. Among these, SA-based mode-locking is preferred to be more environmentally stable and efficient over NPR and NALM [19]. There are different kinds of SAs available such as semiconductor saturable absorber mirrors (SESAMs) [17], carbon nanotubes (CNTs) [18], and graphene [20-21]. Throughout this project, CNT based SAs have been chosen for the experiments to achieve passive mode-locking due to their attractive properties such as the ease of fiber integration, the ultrafast recovery time and the cost-effective production [18, 22-24].

So far SC generation in different nonlinear fibers pumped by CW, nanosecond, picosecond and femtosecond pulse lasers with the relatively long length of fibers have been demonstrated with multi-stage amplification [7-8, 25-26]. In this paper, we demonstrate the SC generation in PCF, HNLF, and ZBLAN fiber by pumping a femtosecond CNT-based passively mode-locked erbium-doped fiber laser (EDFL). A single stage of amplification is applied on passively mode-locked femtosecond pulses to boost the pulse power and to observe the SC spectral variation inside the PCF, HNLF and ZBLAN fiber with respect to the input power variation. The SC spectrum bandwidth of 1050 nm, 1400 nm, and 2000 nm is observed from the output of PCF, HNLF and ZBLAN fiber respectively. In addition to broadband SC generation, the SC bandwidth variations with respect to input power, nonlinear medium, and nonlinear medium length change have also been observed.

2. **Experimental setup**

Fig.1 shows the experimental schematic of SC generation using CNT based passively mode-locked femtosecond EDFL. The 1.55 μm laser ring cavity is built by us in the laboratory. The CNT-SA used in this experiment is with a non-saturable absorption of 27% and a modulation depth of 8%. A 976 nm laser diode pump is coupled to the 0.6 m length of erbium-doped fiber (EDF) through a 980/1550 nm wavelength-division-multiplexing (WDM) coupler. The EDF used in this experiment is a commercial product (LIEKKI Er 110) with a mode-field diameter of around 6.5 $\mu$m and a peak core absorption of 110 dB/m at 1530 nm. An optical isolator is inserted in the cavity to avoid the back reflections and to ensure the optical signal propagates only in one direction. A fiber-based polarization controller (PC) is inserted in order to adjust and optimize the mode-locking condition. 10% of the output is extracted from the cavity using a 90/10 output coupler to analyze the mode-locked pulse characteristics and to use it as a seed laser to the nonlinear fibers. The remaining fibers in the cavity are of standard single-mode fibers and the overall length of the cavity is around 5 meter.

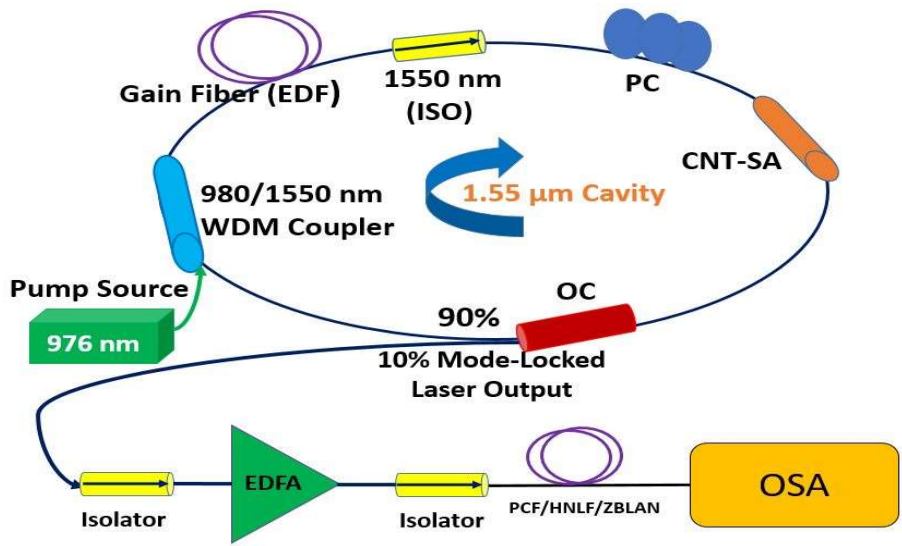

**Fig.1**. Experimental schematic of SC generation using CNT based passively mode-locked femtosecond EDFL, WDM: wavelength division multiplexing, EDF: erbium doped fiber, ISO: isolator, PC: polarization controller, CNT-SA: carbon-nanotube based saturable absorber, OC: output coupler, EDFA: pulsed erbium-doped fiber amplifier, PCF: photonics crystal fiber, HNLF: highly nonlinear fiber, ZBLAN: $ZrF_4$-$BaF_2$-$LaF_3$-$AlF_3$-$NaF$, OSA: Optical spectrum analyzer.

In order to activate the nonlinear phenomena efficiently inside the fibers and also to study the variation of spectral broadening inside the PCF, HNLF and ZBLAN fiber with respect to the variation in input power, the pulse power is boosted using a commercially available pulsed erbium-doped fiber amplifier (EDFA). All the fibers used in this study are commercially available products. Isolators are connected at both the input and the output of the EDFA in order to maintain the stability of the mode-locked cavity and to protect the EDFA from back reflections. The whole setup is maintained throughout the experiment and the output from PCF, HNLF and ZBLAN fiber is measured using an optical spectrum analyzer (OSA). The output SC spectrum generated from each nonlinear fiber is observed using different OSAs covering the spectral measuring range from 450 nm to 3200 nm.

3. **Results and discussion**

*3.1 Characterization of CNT-based passively mode-locked EDFL*

At a certain input pump power, a stable mode-locked pulse is observed at the output of the 1.55 µm cavity. The spectral and temporal characteristics of the mode-locked laser pulses are analyzed using an OSA and a 2 GHz photodetector followed by 350 MHz oscilloscope respectively. Fig.2 depicts the characteristics of the CNT-SA based passively mode-locked EDFL. Fig.2 (a) shows the optical spectrum of the mode-locked pulse at a center wavelength of ~1565 nm with a 3 dB bandwidth of ~5 nm. Fig.2 (b) shows the output pulse train of the mode-locked laser pulse with a pulse repetition rate of 18 MHz. Fig. 2 (c) shows the mode-locked laser pulse auto-correlation trace with sech$^2$ fitting and it has a full width half maximum (FWHM) pulse width of 620 fs.

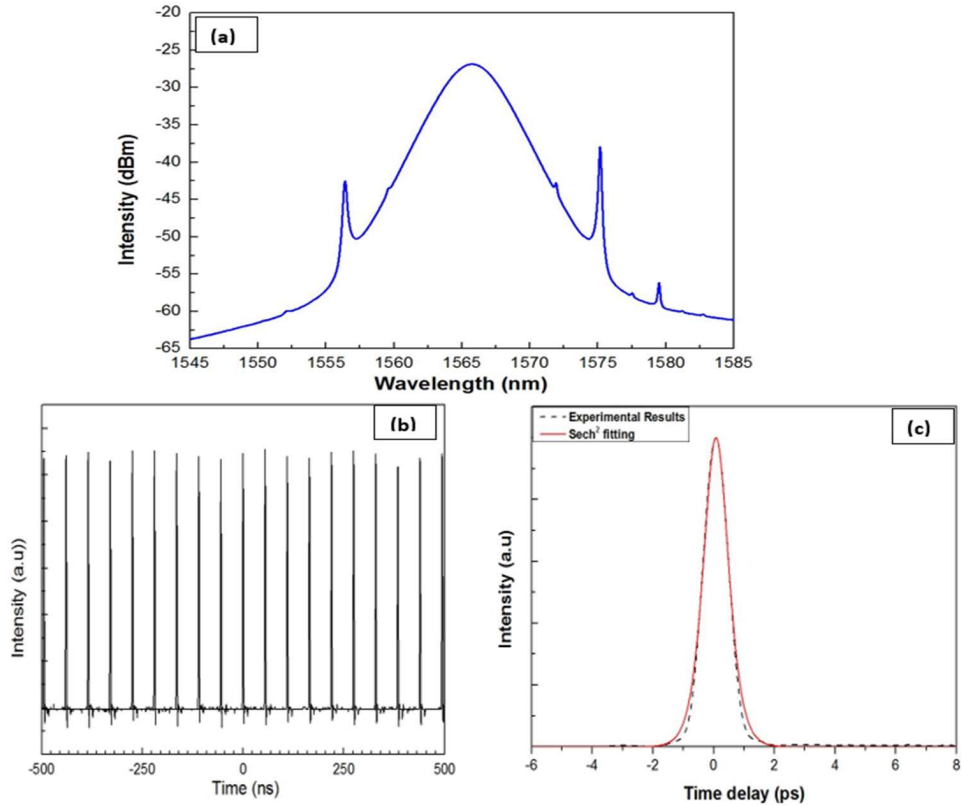

Fig.2 Characterization of CNT-SA based passively mode-locked EDFL: (a) optical spectrum of mode-locked laser pulse, (b) optical pulse train of mode-locked laser pulse, (c) auto-correlation trace of mode-locked laser pulse.

*3.2 Supercontinuum generation using PCF*

Fig. 3 illustrates the variation of output SC spectrum from a 60-meter-long PCF with respect to different input pulse powers. The highly nonlinear PCF has a dispersion of -0.5 ps/ (nm.km), a numerical aperture (NA) of 0.4 and a loss of $\alpha < 9$ dB/km at 1550 nm and a Kerr nonlinearity coefficient of $\gamma > 11$ (W.km)$^{-1}$. It is observed that as the input power to PCF increases, the output spectrum starts broadening on the both sides of the seed pulse wavelength. The femtosecond pulses experiences anomalous dispersion inside the PCF as the seed pulse wavelength (1565 nm) is above the ZDW (<1550 nm) of PCF. Therefore, initially the spectral broadening inside the PCF is dominated by the soliton-fission and self-frequency shift, while as the pulse propagates further inside the fiber, the spectral broadening is strongly assisted by SPM which leads to the symmetrical nature of SC spectrum. A 20 dB bandwidth of 1050 nm SC spectrum extending from 1080 nm to 2130 nm is observed from the output of the 60-meter-long PCF at an input pulse power of 20 dBm. From the observations of spectral broadening phenomena inside the PCF at different input pulse powers, it is understood that at the input pulse powers more than 20 dBm, the spectral broadening is insignificant towards longer wavelength side in our case and the silica-based fibers experience more loss beyond 2 μm due to high material loss. Therefore, in this study, the input pulse power to silica-based nonlinear fibers i.e. PCF and HNLF is limited to 20 dBm and the study is continued by varying the length of the fiber.

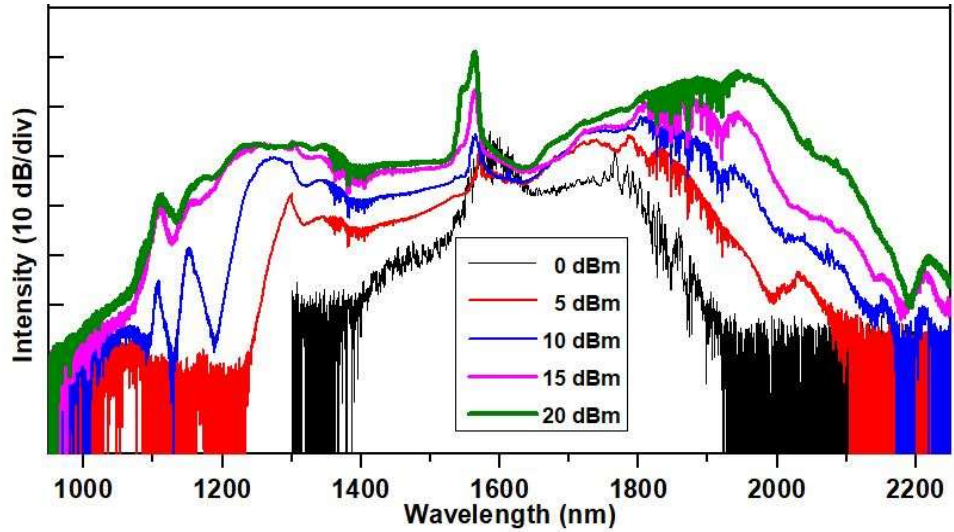

Fig.3 Output SC spectrum from a highly nonlinear PCF (60m) with respect to different input pulse powers.

### 3.3 Supercontinuum generation using HNLF

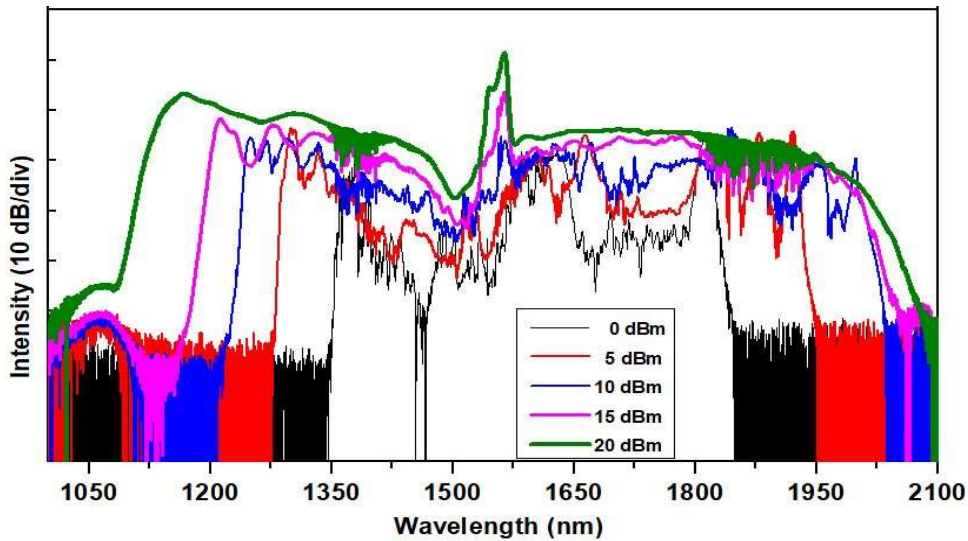

Fig.4 Output SC spectrum from a HNLF (1km) with respect to different input pulse powers.

Fig.4 illustrates the output SC spectrum variation inside the 1-km-long of HNLF with respect to the variation in the input pulse power. The HNLF has a dispersion of -0.23 ps/ (nm.km), a dispersion slope of 0.03 ps/ (nm$^2$.km), loss of α=0.7 dB/km at 1550 nm and Kerr nonlinearity coefficient of γ> 15 (W.km)$^{-1}$. Here also a similar kind of spectral broadening phenomenon is observed as in the case of PCF. The femtosecond pulses experience anomalous dispersion inside the HNLF as the seed pulse wavelength (1565 nm) is above the ZDW (<1550 nm) of HNLF. Therefore, initially, the spectral broadening inside the PCF is dominated by the soliton-fission and self-frequency shift and as the pulse propagates further inside the fiber, the spectral broadening is strongly assisted by SPM which leads to the symmetrical nature of SC spectrum. A 20 dB bandwidth of ~1000 nm SC spectrum (1075 nm - 2075 nm) is observed from the output of the 1-km-long HNLF at an input pulse power of 20 dBm.

After looking at the similar spectral broadening phenomena in both the 60-meter-long PCF (Fig.3) and the 1-km-long HNLF (Fig.4), it is believed that the new frequency generation in any nonlinear medium takes place only in the initial few meters of nonlinear medium. In order to prove this experimentally, we have taken a very short length of HNLF and repeated the experiment. While taking the possible shortest length of HNLF (~1m), the nonlinear length (LNL) is taken into consideration such that the nonlinear phenomena take place efficiently inside the ~1m length of HNLF at all the input pulse powers. $L_{NL}$, is the physical length of the fiber over which nonlinear effects plays the dominant role on the propagating pulse which leads to spectral broadening at the output of the fiber. $L_{NL}= (1/ γP)$, where γ (1/W-km) represents the nonlinearity coefficient of the fiber and P stands for peak power of the input laser pulse. Depending upon the EDFA power level we have calculated the $L_{NL}$, and chosen as ~1m to observe the nonlinear phenomena at all the input pulse powers starting from 0 dBm to 20 dBm.

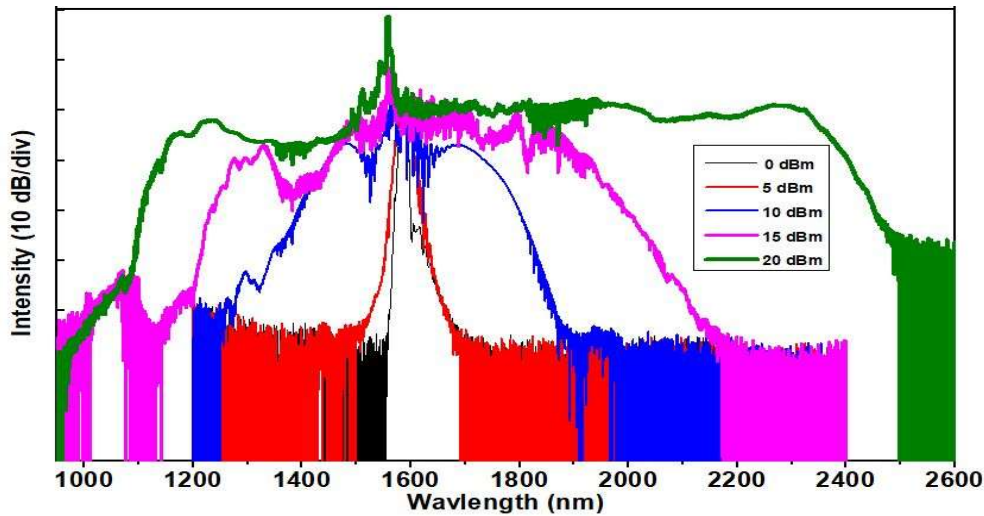

Fig.5 output SC spectrum from a HNLF (<1m) with respect to different input pulse powers.

Fig.5 illustrates the SC spectrum variation inside ~1m length of HNLF with respect to the variation in the input pulse power. As in the case of the 60m length PCF and the 1km length HNLF, it is also observed here that the SC spectrum increases symmetrically on both sides of the input pulse wavelength. A 20 dB bandwidth of ~1400 nm SC spectrum spanning from 1100 nm to 2500 nm is observed inside ~1m length of HNLF at an input power of 20 dBm.

By comparing all the SC output spectrums from 60-meter-long PCF, 1-km-long length HNLF and ~1-meter-long HNLF, it is clear that the nonlinear phenomena are taking place efficiently in the initial few meters' lengths of nonlinear medium providing relatively larger SC bandwidth from the short length of fibers than the long length of fibers. Theory suggests that the longer the fiber length is, the flatter and broader the generated SC can be. However, this statement is only valid only if the nonlinear medium is transparent over the broad wavelength range and the nonlinear medium is of non-dispersive nature. In our case, we have observed the generation of broader SC bandwidth from 1-meter-long HNLF than from the 60-meter-long PCF and 1-km-long HNLF. This is due to two factors: huge material loss of silica-based fibers beyond 2 µm and dispersive nature of the medium. In a dispersive medium, an ultrashort pulse disperses away quickly and vanishes its peak power which leads to the extinction of nonlinear effects even if the pulse propagates further in the nonlinear medium.

### 3.4 Supercontinuum generation using ZBLAN Fiber

Fig.6 illustrates the variation of the output SC spectrum from a 25-meter-long ZBLAN fiber with respect to the variation in input pulse power. The ZBLAN fiber has a core/cladding diameter of 6/125 µm, a NA of 0.26 and a ZDW is at ~ 1.6 µm. The input pulse power to the ZBLAN fiber is varied from 10 dBm to 25 dBm. It is observed that the SC spectrum bandwidth increases with the increase in input pulse power. A 20 dB bandwidth of 2000 nm SC spectrum (1100 nm to 3100 nm) is observed inside the 25-meter-long ZBLAN fiber is observed at an input pulse power of 25 dBm.

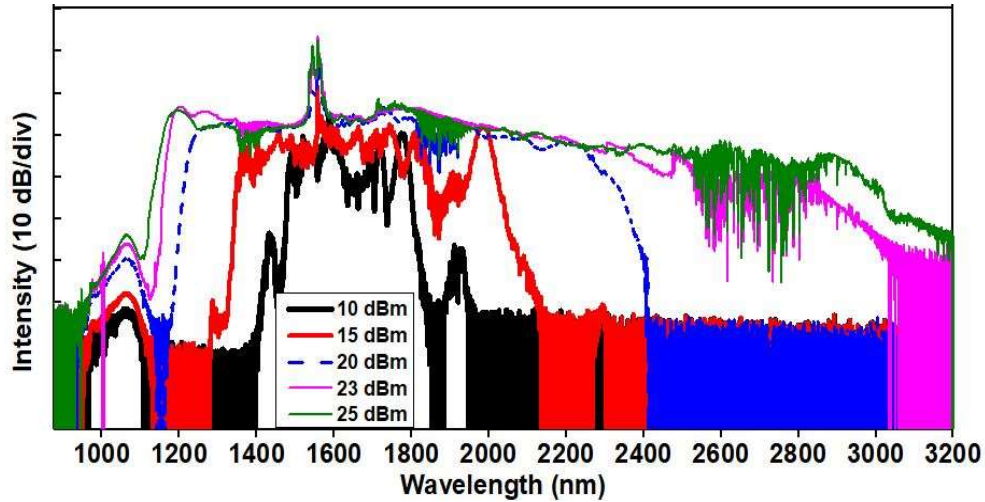

Fig.6 Output SC spectrum from a ZBLAN (25m) with respect to different input pulse powers

To the best of our knowledge, this is the broadest SC spectrum demonstration in ZBLAN fiber using CNT-based passively mode-locked femtosecond EDFL. Here we used only a single stage of amplification and a single stage of SC generation to achieve such a broad SC bandwidth. The femtosecond pulses experience normal dispersion inside the ZBLAN fiber as the seed pulse wavelength (1565 nm) is below the ZDW (1600 nm) of the ZBLAN fiber. Therefore, initially the spectral broadening inside the ZBLAN fiber is dominated by the SPM and soliton-frequency shift, and as the pulse propagates further inside the fiber, the spectral broadening towards longer wavelength side is dominated by the soliton fission. Strong SPM processes enhance the transfer of energy away from pump peak and thereby decrease the residual pump light in the spectrum. Thus, a flatter spectrum is generated by pumping femtosecond pulses in normal dispersion region. The dip that was observed in the spectrum at around 2700 nm corresponds to $OH^{-1}$ ion absorption in the ZBLAN fiber and unpurged detection of long wavelength OSA. Further study of the SC generation inside ZBLAN fiber at higher input powers is limited by the EDFA output power.

## 4. Conclusions

We have demonstrated the broadband SC generation in PCF, HNLF and ZBLAN fiber using CNT-based passively mode-locked EDLF. Passive mode-locking incorporating CNT as SA is achieved a pulse width of 620 fs with a pulse repetition rate of 18 MHZ and a 3 dB bandwidth of ~5 nm at centre wavelength of ~1565 nm. The SC spectrum bandwidth of 1000 nm, 1400 nm, and 2000 nm has been achieved using PCF, HNLF, and ZBLAN fiber respectively and the variation of fiber length and input pulse power is also studied. In this study, we observed the generation of broader SC spectrum from the 1-meter-long HNLF than from the 1-km-long HNLF. This suggests that the short length of silica-based nonlinear fibers is preferable to generate broader SC spectrum at high input pulse powers. A similar kind of spectral broadening phenomenon is observed from the 60-meter-long PCF and 1-km-long HNLF. From the observation of the spectral broadening trend inside the 25-meter-long ZBLAN fiber with respect to the input pulse variation from 10 dBm to 25 dBm, we can conclude that the ZBLAN is one of the best options for mid-IR SC generation. However, in our case, SC spectrum generation further beyond 3.2 μm using ZBLAN fiber is limited by EDFA output power.

## Acknowledgement

The authors would like to acknowledge the technical support by the laboratory managers and technical support officers of Centre for Optical Fibre Technology (COFT) of Nanyang Technological University (NTU), Singapore